\documentclass[aps,prl,twocolumn,reprint,amsmath,amssymb,superscriptaddress,floatfix,footinbib,longbibliography]{revtex4-2}

\usepackage{graphicx}
\usepackage{epstopdf}

\usepackage[T1]{fontenc}
\usepackage[applemac]{inputenc}
\usepackage{lmodern}
\usepackage{amsmath}
\usepackage{amssymb}
\usepackage[english]{babel}
\usepackage{natbib}
\usepackage{ae}
\usepackage{units}
\usepackage{ulem}
\usepackage{amsmath,amssymb,natbib,bm}
\usepackage{psfrag}
\usepackage{subfigure}
\usepackage{amsthm}

\usepackage{booktabs}
\usepackage{slashed}

\usepackage[americaninductors]{circuitikz}
\usepackage{tikz}
\usetikzlibrary{arrows}
\usepackage{ulem}
\usepackage{color}
\usepackage{url}

\usepackage[colorlinks]{hyperref}
\hypersetup{%
        plainpages=true,
        breaklinks=true,
        hypertexnames=false,
        pageanchor=true,
        colorlinks=true,
        linkcolor={blue},
        citecolor={magenta},
        urlcolor={blue},
        anchorcolor={black}
      }

\usepackage{mleftright} 

\newcommand{\figref}[1]{\mbox{Fig.~\ref{#1}}}

\renewcommand{\eqref}[1]{\mbox{Eq.~(\ref{#1})}}
\newcommand{\figpanel}[2]{Fig.~\hyperref[#1]{\ref*{#1}(#2)}} 

\newcommand{\figpanels}[3]{Figs.~\hyperref[#1]{\ref*{#1}(#2,#3)}} 
\newcommand{\figpanelNoPrefix}[2]{\hyperref[#1]{\ref*{#1}(#2)}} 
\newcommand{\figpanelsNoPrefix}[3]{\hyperref[#1]{\ref*{#1}(#2)-(#3)}} 

\newcommand{\be}{\begin{equation}}
\newcommand{\ee}{\end{equation}}
\newcommand{\bea}{\begin{eqnarray}}
\newcommand{\eea}{\end{eqnarray}}

\AtBeginDocument{%
    \newwrite\bibnotes
    \def\bibnotesext{Notes.bib}
    \immediate\openout\bibnotes=\jobname\bibnotesext
    \immediate\write\bibnotes{@CONTROL{REVTEX42Control}}
    \immediate\write\bibnotes{@CONTROL{%
    apsrev42Control,author="08",editor="1",pages="0",title="0",year="1"}}
     \if@filesw
     \immediate\write\@auxout{\string\citation{apsrev42Control}}%
    \fi
}%

\begin{document}

\title{Microwave interference from a spin ensemble and its mirror image in waveguide magnonics} 

\author{B.-Y.~Wu}
\thanks{These authors contributed equally}
\affiliation{Department of Physics, City University of Hong Kong, Kowloon, Hong Kong SAR, China}

\author{Y.-T.~Cheng}
\thanks{These authors contributed equally}
\affiliation{Department of Physics, City University of Hong Kong, Kowloon, Hong Kong SAR, China} 

\author{K.-T.~Lin}
\thanks{These authors contributed equally}
\affiliation{Department of Physics and Center for Quantum Science and Engineering, National Taiwan University, Taipei 10617, Taiwan}

\author{F. Aziz}
\affiliation{Department of Physics, City University of Hong Kong, Kowloon, Hong Kong SAR, China}
\affiliation{Department of Physics, National Tsing Hua University, Hsinchu 30013, Taiwan}

\author{J.-C.~Liu}
\affiliation{Department of Electronic and Computer Engineering, The Hong Kong University of Science and Technology, Clear Water Bay, Kowloon, Hong Kong SAR, China}

\author{K.-V. Rangdhol}
\affiliation{Department of Physics, City University of Hong Kong, Kowloon, Hong Kong SAR, China}

\author{Y.-Y. Yeung}
\affiliation{Department of Physics, City University of Hong Kong, Kowloon, Hong Kong SAR, China}

\author{Sen~Yang}
\affiliation{Department of Physics and IAS Center of Quantum Technologies, The Hong Kong University of Science and Technology, Clear Water Bay, Kowloon, Hong Kong SAR, China}

\author{Qiming~Shao}
\affiliation{Department of Electronic and Computer Engineering, The Hong Kong University of Science and Technology, Clear Water Bay, Kowloon, Hong Kong SAR, China}

\author{Xin Wang}
\affiliation{Department of Physics, City University of Hong Kong, Kowloon, Hong Kong SAR, China}

\author{G.-D.~Lin}
\affiliation{Department of Physics and Center for Quantum Science and Engineering, National Taiwan University, Taipei 10617, Taiwan}
\affiliation{Physics Division, National Center for Theoretical Sciences, Taipei 10617, Taiwan}
\affiliation{Trapped-Ion Quantum Computing Laboratory, Hon Hai Research Institute, Taipei 11492, Taiwan}

\author{Franco Nori}
\affiliation{Center for Quantum computing, and Cluster for Pioneering Research, RIKEN, Wakoshi, Saitama 351-0198, Japan}
\affiliation
{\mbox{physics Department, The University of Michigan, Ann Arbor, Michigan 48109-1040, United States}}

\author{I.-C.~Hoi}
\email[e-mail:]{iochoi@cityu.edu.hk}
\affiliation{Department of Physics, City University of Hong Kong, Kowloon, Hong Kong SAR, China}

\date{\today}
\begin{abstract}
We investigate microwave interference from a spin ensemble and its mirror image in a one-dimensional waveguide. 
Away from the mirror, the resonance frequencies of the Kittel mode (KM) inside a ferrimagnetic spin ensemble have sinusoidal shifts as the normalized distance between the spin ensemble and the mirror increases compared to the setup without the mirror.
These shifts are a consequence of the KM's interaction with its own image.
Furthermore, the variation of the magnon radiative decay into the waveguide shows a cosine squared oscillation and is enhanced twofold when the KM sits at the magnetic antinode of the corresponding eigenmode.  
We can finely tune the KM to achieve the maximum adsorption of the input photons at the critical coupling point.
Moreover, by placing the KM in proximity to the node of the resonance field, its \textit{lifetime is extended to more than eight times} compared to its positioning near the antinode.

\end{abstract}

\maketitle
\textit{Introduction.}---Recent studies on spin ensembles have utilized light-matter interaction~\cite{Scully_Zubairy_1997,Agarwal_2012} through collective effects for coherent information processing in both the quantum and classical regimes~\cite{Lukin2003,Atac2009}.
Significant advances have been made in cavity magnonics~\cite{ZARERAMESHTI20221}, demonstrating (ultra) strong coupling between a magnon mode and a cavity mode~\cite{Tabuchi2014,Zhang2014,Goryachev_2014,Zhang2015,Bourhill2016, Kostylev2016,Goryachev_2018,Flower_2019}, information transduction~\cite{Xufengscienceadvance2016,Hisatomi2016,Osada2016}, bistability~\cite{Yi-Pu2018,Nair_2021,Shen2022}, exceptional points~\cite{ZhangNC2017,ZhangPRB2019}, memory applications~\citep{Zhang2015NC,ShenPRL2021}, and proposing the generation of entangled quantum states between magnons, cavity photons and phonons~\cite{Li_PRL_2018,Zhangzhedong_2019}, as well as magnon squeezed states~\cite{Li_PRA_2019}.
Additionally, coupling between magnons and a superconducting qubit, mediated by cavity photons, has been achieved~\cite{TabuchiScience2015,Danysciadv2017, DanyScience2020,XuPRL2023}. 
Beyond single modes, coherence transfer through continuous photonic modes have been investigated in waveguide magnonics~\cite{Rao2020,Wang2022,Qian2023,Wang_NP_2024}, opening up avenues for studying giant atom physics~\cite{Wang2022} and developing non-Hermitian physics~\citep{Qian2023}.

Since Purcell's seminal work~\citep{Purcell1946}, the implementation of a simple ``cavity" boundary with a single mirror has supported a fully open system with continuous photonic modes. 
This configuration allows for the engineering of electromagnetic field modes around an atom~\cite{Buluta_2009,Georgescu_RMP_2014}, thereby modifying its emission properties through mirror-induced interference.
For example, the fluorescence of a single $\rm Ba^{+}$ ion near a floating mirror is either enhanced or reduced due to the interaction between the ion and its mirror image via optically radiative means~\citep{Eschner2001}.
In another instance, a superconducting artificial atom (transmon)-mirror system demonstrated that, using a mirror, the device enables the detection of the spectral density of vacuum fluctuations~\citep{Hoi2015}, the study of Landau-Zener-St{\"u}ckelberg-Majorana interferometry~\citep{SHEVCHENKO20101_PR,Wen_PRB_2020,Liul_PRB_2023}, and the deterministic loading of microwaves~\citep{Lin2022}.  
More importantly, mirror-shaped photonic modes can shift energy levels through virtual photon processes~\cite{Meschede1990,Dorner2002}.
For example, in a three-dimensional (3D) single atom far from a mirror~\cite{Wilson_2003}, the level shifts are approximately of $\pm\unit[150]{kHz}$, a value that is limited due to the small effective solid angle.
In a 1D geometry, implemented by coupling the artificial atom to a semi-infinite waveguide, theoretical predictions show that the energy shifts vary periodically through effectively altering the atom-mirror distance~\cite{Koshino_2012}. 
However, such an energy level shift and its relation to radiative decay have not been observed in this 1D setup due to the absence of a reference setup~\citep{Hoi2015}. 
Additionally, controlling the atom's lifetime in such configuration in the time domain is still unexplored.

\begin{figure}
\includegraphics[width=\linewidth]{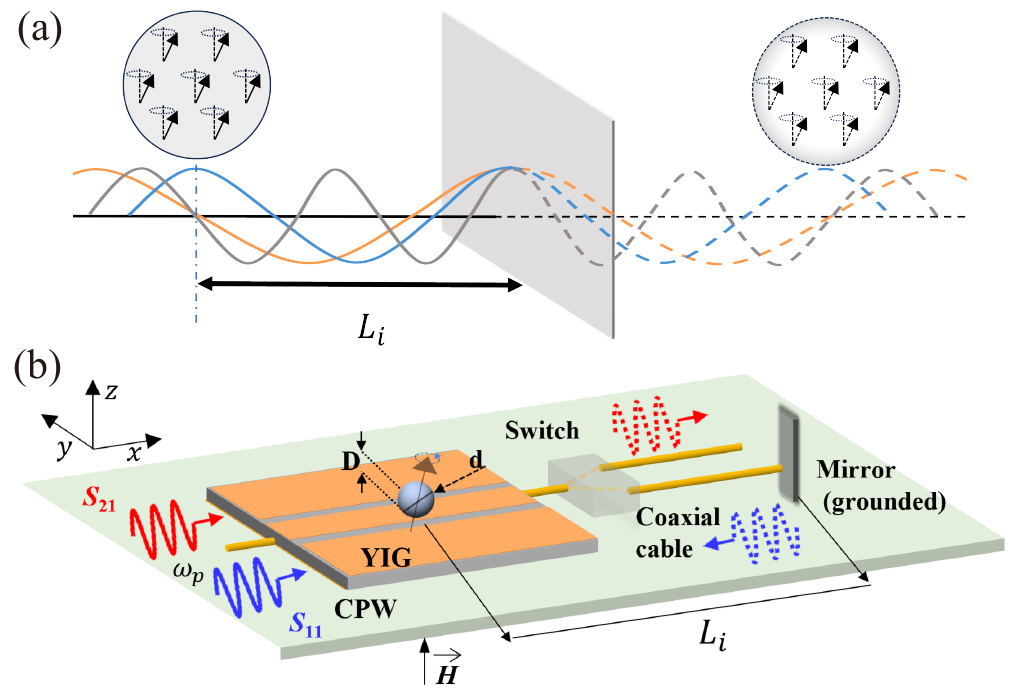}
\caption{
Schematic of the system and experimental setup.
(a) A YIG sphere is positioned at a distance $L_{i}$ from the mirror.
The three color curves show the mode structures of the propagating magnetic field. 
The coupling between the magnetic field (orange) and the KM is minimal when the KM sits at the node (vertical dashed line) of the propagating resonant field.
(b) The experimental setup is composed of a YIG sphere with a diameter $d$ being placed above the CPW plane at a height $D$. 
In the grounded mirror system, a short load is used as the mirror to reflect the microwave at the end of the path.  
By measuring the reflection coefficient $S_{11}$ (blue waves), the mirror-induced interference effects can be probed and analyzed.
The system without the mirror is considered as a reference, characterized by measuring the transmission coefficient $S_{21}$ (red waves).
}
\label{fig:setup}
\end{figure}

In this work, we use a Yttrium Iron Garnet (YIG, $\rm Y_{3}Fe_{5}O_{12}$) sphere as an ideal spin ensemble with a high spin density $\unit[2.1\times10^{22}\mu_{B}]{cm^{-3}}$~($\mu_B$:~Bohr magneton)~\cite{Gilleo1958, Huebl2013} and low intrinsic loss~\citep{Zhang2014,Tabuchi2014}, to explore \textit{mirror induced interference effects}. 
By applying an external magnetic field, the magnon modes (dipolar spin waves or magnetostatic modes) of this YIG can be detected through microwave excitation.
These magnon modes, arising from the collective behaviour of electron spins, exhibiting bosonic characteristics~\cite{Daniel2009}.
In this work, we focus on investigating the interference between a particular magnon mode, the Kittel mode (KM), and its mirror image in a 1D geometry [see~\figpanel{fig:setup}{a}], where the precession of spins is in phase~\cite{Dicke1954}.
This setup~\citep{Astafiev2010,You2011,Hoi2011, Hoi2012,Hoi2013,GU20171,Kockum_2019,Wen_PRL_2018,Wen2019} \textit{enhances interactions and reduces the decay into unwanted modes compared to the 3D cases}~\citep{Eschner2001}.
To observe the KM resonance shifts, we construct two systems, one with a grounded mirror and one without, which can be selected by a microwave switch [see~\figpanel{fig:setup}{b}].
By tuning the KM frequency, we demonstrate that the continuous resonance shifts vary periodically over a full wavelength.
The maximum shift depends on the round trip phase and the YIG size. 
Additionally, with the mirror, the excited KM interferes with its own radiation, causing a change in the radiative decay, which is enhanced at the antinodes and diminished at nodes under the corresponding eigenmodes, thereby altering the KM's lifetime. 
This modification enables the formation of \textit{controllable ultra-sharp adsorption dips}.

\textit{System and model.}---Our experimental setup, shown in ~\figpanel{fig:setup}{b}, consists of a YIG sphere placed on top of a coplanar waveguide (CPW). 
We can extensively tune the KM or other magnon modes at room temperature.
The KM, regarded as having a large magnetic dipole moment~\citep{Zhang2014,Tabuchi2014}, enables strong magnon-photon coupling.
In the system with a mirror, the coherent input interacts with the KM and continues propagating towards the mirror.
When the KM is excited, the emitted field is evenly distributed between the left- and right-moving fields. 
Both the input and the right-moving field are reflected by the mirror and subsequently interfere with the left-moving field.
Notably, a round trip time or a delay time is much smaller ($2L_{i}/\nu \approx\unit[2\sim5]{ns}$) compared to the lifetime of the KM ($\unit[\approx70]{ns}$).
Therefore, the magnon-photon interaction for this feedback-included system occurs in a Markovian regime~\citep{Crowder2022}.
Here, $L_{i}\approx21\sim53\:{\rm cm}$ represents the distance between the YIG sphere and the mirror, where in this work we have four different distances ($i=1,2,3,4$), and $\nu\approx \unit[2.1\times10^{8}]{m/s}$ denotes the speed of the light~(Section S2 in \cite{SupMat}).

The phase resulting from the round trip is given by (Eq.~(S9)~\cite{SupMat})
\begin{equation}
	{\theta(I)} = 2\times2\pi L_{i} /\lambda(I),
	\label{Eq:phase}
\end{equation}
where $\lambda(I)=2\pi\nu/\omega_{m,t}(I)$ is the wavelength of the KM transition frequency $\omega_{m,t}$ without considering the mirror effect and $I$ denotes the current passing through the electromagnet. 
The subscripts $t$ (transmission) and $r$ (reflection) refer to the cases without and with the mirror, respectively.
Furthermore, due to the presence of the mirror, the radiative damping rate $\kappa_{r}$ and the resonant shift $\delta\omega\equiv\omega_{m,r}-\omega_{m,t}$
can be related to the phase $\theta\left(I\right)$ through:
\begin{equation}
	{\kappa_{r}}=(2\kappa_{b})\cos^2{[\theta(I)/2]}
	\label{Eq:rediative_damping_rate}
\end{equation}
and 
\begin{equation}
   \delta\omega={(\kappa_b/2)}\sin[\theta(I)], \
   	\label{Eq:resonance_shift}
\end{equation}
respectively (Eqs.~(S7,S8)~\cite{SupMat}). 
Here $\kappa_{b}$ denotes the bare radiative damping rate.

\begin{figure*}[t!]
\includegraphics[width=\linewidth]{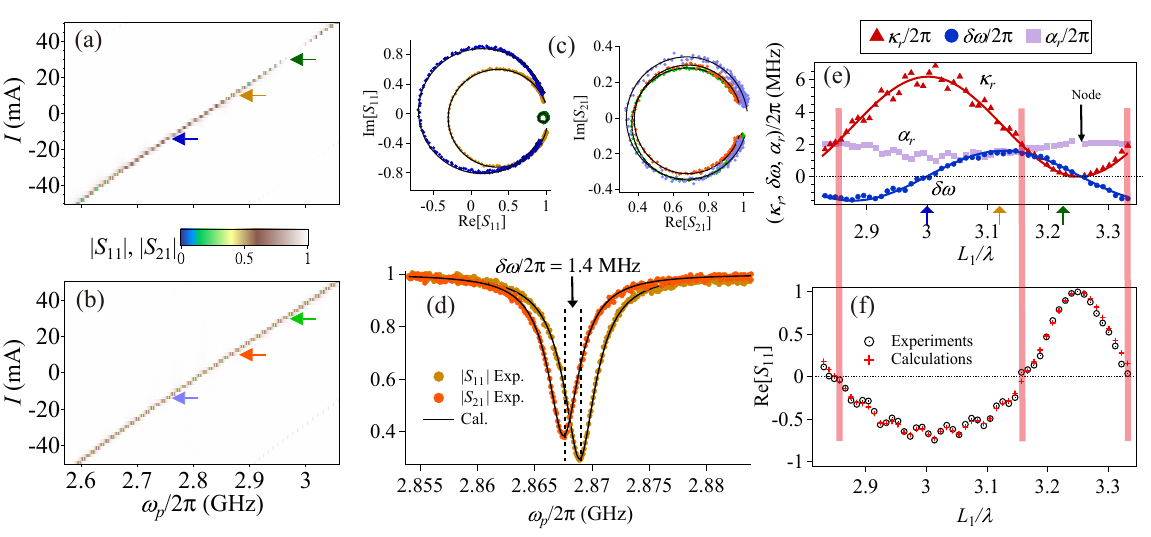}
\caption{
Spectroscopy of the YIG sphere with diameter $d=\unit[1.0]{mm}$. 
(a) Reflection spectrum $|S_{11}|$ versus the current $I$ for the YIG-mirror system.
(b) Transmission spectrum $|S_{21}|$ against the current $I$ for the system \textit{without} the mirror.
In addition to the Kittel mode to be probed, a high-order magnetostatic mode is observed in the lower right corner of (a) and (b).
(c) In-phase-and-Quadrature (IQ) plots for the reflection and transmission spectra. 
The left and right panels display complex plane representations of the reflection and transmission coefficients ($[S_{11}]$ and $[S_{21}]$) for three different resonance frequencies of this YIG, as indicated by the horizontal arrows in (a).
In this representation, the coefficient $[S_{11}]$ and $[S_{21}]$ form circles, with the diameter of these circles corresponding to the KM's linewidth $\Gamma_{r\left(t\right)}$.
Without special instructions, symbols are experimental data.
Here, the solid curves follow Eqs.~(\ref{Eq:reflection_coefficient},\ref{Eq:tansmission_coefficient}).
The extracted parameters are given in~Table.~\ref{tab:Parameter1}.
(d) The line-cut plot for both reflection and transmission spectra for the current $I=\unit[8]{mA}$.
(e) The radiative damping rate $\kappa_{r}/2\pi$ (red triangles), resonant shifts $\delta\omega/2\pi$ (blue dots), and non-radiative damping rate $\alpha_{r}/2\pi$ (purple squares) versus the normalized distance, $L_{1}/\lambda$, with $\lambda=\nu/(\omega_{m.t}/2\pi)$ being the resonant wavelength, for the KM of a YIG away from the mirror. 
Here, the red (blue) solid curve is based on \eqref{Eq:rediative_damping_rate} [\eqref{Eq:resonance_shift}].
The short vertical arrows on the bottom axis in (e) correspond to the three horizontal arrows shown in (a). 
(f) Real part of the reflection $[S_{11}]$ as a function of the effective distance $L_{1}/\lambda$.
The critical coupling points are indicated by the red faint lines, which are formed when $\kappa_{r}=\alpha_{r}$, and $\omega_{p}=\omega\rm_{m,r}$.
 }
\label{fig:fig2}
\end{figure*}
\begin{table*}[t!]
	\centering
	\begin{tabular}{| c | c | c | c | c | c | c | c | c | c |}
		\hline
		&
		$\omega_{m,r} / 2 \pi$&
		$\kappa_{r} / 2 \pi$&
		$\Gamma_{r} / 2 \pi$&
            $\alpha_{r} / 2 \pi$&
		&
		$\omega_{m,t} / 2 \pi$&
		$\kappa_{t} / 2 \pi$&
		$\Gamma_{t} / 2 \pi$&
            $\alpha_{t} / 2 \pi$\\
		\hline
		&
		GHz&
		MHz&
		MHz&
            MHz&
		&
		GHz&
		MHz&
		MHz&
            MHz\\
		\hline
		\definecolor{myblue}{RGB}{0,0,255}
		\tikz[baseline=(current bounding box.center)] \draw[-triangle 45, myblue, thick] (0,0) -- (1,0);&
		$2.7549$ &
		$6.434$ &
		$3.797$ &
            $1.160$ &
		\definecolor{myblue2}{RGB}{153,153,255}
		\tikz[baseline=(current bounding box.center)] \draw[-triangle 45, myblue2, thick] (0,0) -- (1,0);&
		$2.7551$ &
		$3.291$ &
		$2.438$ &
            $1.585$ \\
		\hline
		\definecolor{myyellow2}{RGB}{203,135,0}
		\tikz[baseline=(current bounding box.center)] \draw[-triangle 45, myyellow2, thick] (0,0) -- (1,0);&
		$2.8689$ &
		$2.969$ &
		$2.291$ &
            $1.613$ &
            \definecolor{myyellow3}{RGB}{255,85,0}
		\tikz[baseline=(current bounding box.center)] \draw[-triangle 45, myyellow3, thick] (0,0) -- (1,0);&
		$2.8675$ &
		$2.614$ &
		$2.124$ &
            $1.634$ \\
		\hline
		\definecolor{mygreen2}{RGB}{0,128,0}
		\tikz[baseline=(current bounding box.center)] \draw[-triangle 45, mygreen2, thick] (0,0) -- (1,0);&
		$2.9628$ &
		$0.131$ &
		$1.070$ &
            $2.009$ &
		\tikz[baseline=(current bounding box.center)] \draw[-triangle 45, green, thick] (0,0) -- (1,0);&
		$2.9623$ &
		$3.063$ &
		$2.442$ &
            $1.821$ \\
		\hline
	\end{tabular}
	\caption{Summary of the Kittel mode parameters indicated by the six horizontal color arrows in~\figpanels{fig:fig2}{a}{b}.}
	\label{tab:Parameter1}
\end{table*}


\textit{Results and discussions.}---To investigate the interference effects, we adjust the distance $L_{i}$ by varying the length of the coaxial cable. 
Initially, we set $L_{1}=\unit[22.8]{cm}$ (see section S2~\citep{SupMat}) and use a YIG sphere with a diameter $d=\unit[1.0]{mm}$.
We introduce a tunable static magnetic field $\vec{H}$ adjusted by the applied current $I$, which is perpendicular to the CPW plane and to the YIG's crystal axis <110>, to control the magnon resonance frequency [see~\figpanel{fig:setup}{b}].
The sample is assumed to be uniformly magnetized.
We characterize the system by measuring the reflection coefficient $S_{11}$ in the mirror setup and the transmission coefficient $S_{21}$ in the reference setup without the mirror.
The probe power is set to $\unit[-30]{dBm}$ to ensure the KM's excitation remained in the linear regime.
In the steady state, the reflection and transmission coefficients are given by~(Eqs.~(S11,S20)~\cite{SupMat}):
\begin{equation}
	{S\rm_{11}} =   1 - \frac{\kappa_{r}}{{\Gamma_{r}- i{(\omega_p-\omega_{m,r})}}}
	\label{Eq:reflection_coefficient}
\end{equation}
and
\begin{equation}
	{S\rm_{21}} =   1 - \frac{\kappa_{t}/2}{{\Gamma_{t}- i{(\omega_p-\omega_{m,t})}}},
	\label{Eq:tansmission_coefficient}
\end{equation}
where $\Gamma_{r(t)}=(\kappa_{r(t)} + \alpha_{r(t)})/2$, with $\alpha_{r(t)}$ denoting the non-radiative decay, represents the overall damping rate (linewidth) of the KM. 

In~\figpanels{fig:fig2}{a}{b} we plot the reflection and transmission spectra as a function of the applied current $I$.
We find that the absorption dips occur at the probe frequencies $\omega_{p}$ equal to
the KM resonance frequencies $\omega_{m,r\left(t\right)}$, allowing us to here focus only on the KM. 
When $I=0$, the external permanent magnet and the internal anisotropy field of the YIG sphere contribute to $\omega_{m,t}/2\pi=\unit[2.8204]{GHz}$.
To analyze the KM behavior, we examine the spectroscopic line shape.
Near the resonant dips indicated by arrows in~\figpanels{fig:fig2}{a}{b}, the complex plane representations of $S_{21}$ and $S_{11}$ manifest themselves as circles~\citep{Probst2015}, as shown in~\figpanel{fig:fig2}{c}, with the diameter of these circles corresponding to the KM's linewidth $\Gamma_{r(t)}$.
Smaller diameters indicate weaker magnon-photon coupling.
By using Eqs.~(\ref{Eq:reflection_coefficient},\ref{Eq:tansmission_coefficient}), we can extract $\kappa_{r(t)}, \Gamma_{r(t)}, \omega_{m, r(t)}$, and $\alpha_{r(t)}$ (see Table.~\ref{tab:Parameter1}).  
When $\omega_{m,r}/2\pi$ is around $\unit[2.9628]{GHz}$, the coupling strength approaches zero, inhibiting real photon exchange~\citep{Kockum2018,Kannan2020}.
This occurs when the KM is located at the magnetic node of the corresponding eigenmode, effectively concealing itself from the probe field.  
In the reference setup, the radiative damping $\kappa_{t}$ remains in the same order of magnitude ($\kappa_{t}/2\pi\approx\unit[3]{MHz}$) for these three cases.  
 
We then show the extracted $\kappa_{r}/2\pi$ (red triangles) versus distance $L_{1}/\lambda$ in~\figpanel{fig:fig2}{e} by fitting all spectra shown in~\figpanel{fig:fig2}{a}.
We first observe that the theoretical calculation, using Eqs.~(\ref{Eq:phase},\ref{Eq:rediative_damping_rate}), aligns well with our measurement, allowing us to extract the bare damping rate $\kappa_{b}/2\pi=\unit[3.092]{MHz}$.
When the normalized distance is near $L_{1}/\lambda=3.25$, $\kappa_{r}/2\pi$ approaches zero, signaling that the KM is at the node of its eigenmode. 
However, when $L_{1}/\lambda=3$, we observe $\kappa_{r}/2\pi=\unit[6.434]{MHz}$ which is nearly twice that of the same $I$, where $\kappa_{t}/2\pi=\unit[3.291]{MHz}$, suggesting that the waveguide-magnon coupling is enhanced by putting the YIG at the antinode of its eigenmode.

Furthermore, the interaction between the KM and its mirror image causes energy-level shifts.
We then demonstrate this from the reflection and transmission spectra when $I=\unit[8]{mA}$, as shown in~\figpanel{fig:fig2}{d}.
We observe that the resonance frequency shift $\delta\omega=\omega_{m,r}-\omega_{m,t}=2\pi\times1.4\:{\rm MHz}$, after correcting for impedance mismatch~\citep{Probst2015}.
To further study this shift, we measure the resonance frequency shift $\delta\omega$ against the distance $L_{1}/\lambda$, as depicted in~\figpanel{fig:fig2}{e}.
Our theoretical calculations, using Eqs.~(\ref{Eq:phase},\ref{Eq:resonance_shift}), fit the experimental data well, validating the \textit{self-interaction} of the YIG and its image, resulting in peak-to-peak frequency shifts of approximately $\kappa_b/2=2\pi\times\unit[\pm 1.54]{MHz}$.
Interestingly, when $L_{1}/\lambda=3$, the resonance frequency shift vanishes ($\delta\omega=0$), even with maximum magnon-photon coupling. 
This is attributed to the accumulated phase $\theta(I)$ at the antinode being an integer multiple of $\pi$, according to Eqs.~(\ref{Eq:phase},\ref{Eq:rediative_damping_rate},\ref{Eq:resonance_shift}).
We also notice that the maximum resonant shift occurs at the midpoint between the antinode and node, and the position-dependent shift can be either negative or positive.

In~\figpanel{fig:fig2}{f} we show the real part of the reflection coefficient
$S_{11}$ versus the distance $L_{1}/\lambda$ in the resonant case
when $\omega_{p}=\omega_{m,r}$.
Note that, in this plot, the imaginary part of $S_{11}$ is close to zero.
Negative Re$[S_{11}]$, resulting from the $\pi$ phase shift in the output voltage, indicates strong coupling between the YIG and resonant microwave. 
According to \eqref{Eq:reflection_coefficient} we find that the value of Re[$S_{11}$] depends on the magnitude ratio between $\kappa_{r}$ and $\Gamma_{r}$. 
We observe Re$[S_{11}]=0$, indicative of all the coherent input photons that lose their energy, forming an ultra-sharp adsorption dip [see an example in~Fig.~S4(a)~\cite{SupMat} that reaches $\unit[-49.7]{dB}$].
This satisfies the critical coupling condition~\cite{Gorodetsky1999,Cai2000,Koshino_2012}, requiring $\kappa_{r}=\alpha_{r}$, indicated by the red faint lines connecting~Figs.~\ref{fig:fig2}(e) and~\ref{fig:fig2}(f). 
By adjusting the distance from $L_{2}$ ($\unit[38.1]{cm}$) to $L_{3}$ ($\unit[53.3]{cm}$), the critical coupling points are shifted and increased [see Section S3 \citep{SupMat}], aligning with the predicted periods of $\kappa_{r}$ and $\delta\omega$.
This observation validates the theoretical predictions through Eqs.~(\ref{Eq:rediative_damping_rate},\ref{Eq:resonance_shift}).

\begin{figure}[ht]
\centering
\includegraphics[width=\linewidth]{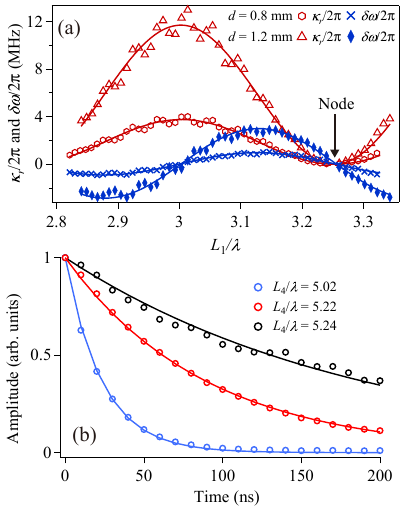}
\caption{Collective coupling and temporal dynamic emission.
(a) The variations of the radiative damping rate $\kappa_{r}/2\pi$ and the resonance shifts $\delta\omega/2\pi$ of the KM with the effective distance $L_{1}/\lambda$ for YIG spheres with different diameters $d=\unit[0.8, 1.2]{mm}$. 
Here, symbols are experimental data that are obtained from the same setup in~\figref{fig:setup}.   
The red (blue) theoretical curves in (a) is based on \eqref{Eq:rediative_damping_rate} (\eqref{Eq:resonance_shift}) with $\kappa_{b}/2\pi=\unit[5.845]{MHz}$ for $d=\unit[1.2]{mm}$ ($\kappa_{b}/2\pi=\unit[1.946]{MHz}$ for $d=\unit[0.8]{mm}$).
(b) The temporal dynamics of the excited KM of the YIG sphere with $d=\unit[1.2]{mm}$ when it sits at $L_{4}/\lambda$ [see Fig. S5~\citep{SupMat} for details].
The circles represent the experimental data.
The solid curves are fitted by the exponential function, which gives lifetimes: $\unit[23.1\pm0.1]{ns}$ (blue), $\unit[89.0\pm0.2]{ns}$ (red), and $\unit[188.4\pm12.1]{ns}$ (black). 
The experimental setup is shown in Fig.~S1(b)~\citep{SupMat}.
}
\label{fig:gap_distance_time_domain}
\end{figure}
	
We now turn to investigate the dependence of $\kappa_{r}$
and $\delta\omega$ on the size of the
YIG sphere, which directly correlates with the number of spins. 
Figure~\ref{fig:gap_distance_time_domain}(a) shows an enhancement in the interference fringe amplitude for the large sphere ($d=\unit[1.2]{mm}$) with $\kappa_b/2\pi=\unit[5.845]{MHz}$ (see Table.~S1 for details~\cite{SupMat}).
This proves that increasing the number of spins leads to an \textit{enhanced collective effect and a stronger resonant dipole-dipole interaction with its mirror image}.
Moreover, the spatial distribution of the field in the CPW also affects $\kappa_{t}$~\citep{Rao_APL_2017,Yang2018}.
When the height $D$ (see~\figref{fig:setup}) is larger than $d$, $\kappa_{t}$ is proportional to the number of spins (see Fig.~S6(b)~\cite{SupMat}).
We further characterize the grounded mirror system in the time domain with a YIG sphere of $d=\unit[1.2]{mm}$ using a square pulse excitation.
After the resonant pulse excites the KM, an exponential decay of the photon energy is observed, as shown in~\figpanel{fig:gap_distance_time_domain}{b}.
For the case where $L_{4}/\lambda=5.02$ and $\omega_{m,r}/2\pi=\unit[3.008]{GHz}$, the magnon lifetime $\tau=\unit[23.1\pm0.1]{ns}$ is obtained by directly fitting the experimental data with an exponential function, in agreement with the frequency domain result via $\tau=1/\Gamma_{r}=\unit[23.0]{ns}$ with $\Gamma_{r}/2\pi=\unit[6.919]{MHz}$.
The lifetime of the KM located near the node ($L_{4}/\lambda=5.24$) is $\tau=\unit[188.4\pm12.1]{ns}$, although its output emission is very weak [see Fig.~S5(b)~\cite{SupMat}].
\textit{The lifetime near the node is more than eight times longer than that near the antinode}.
These results demonstrate our ability to \textit{finely control the magnon lifetime, leading to either reduced or enhanced emission}, all achieved without physically moving the sphere.

\textit{Summary and outlook.}---In the mirror-embedded 1D waveguide magnonic system, the KM exhibits sinusoidal resonant shifts and radiative damping oscillations due to its interference with its own radiation.
These results arise from the coupling to continuous photonic modes.
Consequently, we are empowered to perform precise and continuous tuning of the real photon exchange processes, ranging from maximal coupling to decoupling between the KM and photons.
This gives us control over the formation of critical couplings, characterized by the perfect adsorption.
Finally, the ability to control the lifetime of magnons provides a valuable tool for coherent information processing.


\begin{acknowledgments}
\textit{Acknowledgements.}---I.-C.H.~acknowledges financial support from City University of Hong Kong through the start-up project 9610569, from the Research Grants Council of Hong Kong (Grant No.~11312322) and from Guangdong Provincial Quantum Science Strategic Initiative (GDZX2303005 and GDZX2200001).
F.N.~is supported in part by: Nippon Telegraph and Telephone Corporation (NTT) Research, the Japan Science and Technology Agency (JST) [via the Quantum Leap Flagship Program (Q-LEAP), and the Moonshot R$\&$D Grant Number JPMJMS2061], and the Office of Naval Research (ONR) Global (via Grant No. N62909-23-1-2074).

\end{acknowledgments}

\normalem
\bibliography{References}

\end{document}


\title{Supplementary Material for ``Microwave interference from a spin ensemble and its mirror image in waveguide magnonics"}

\author{B.-Y.~Wu}
\thanks{These authors contributed equally}
\affiliation{Department of Physics, City University of Hong Kong, Kowloon, Hong Kong SAR, China}

\author{Y.-T.~Cheng}
\thanks{These authors contributed equally}
\affiliation{Department of Physics, City University of Hong Kong, Kowloon, Hong Kong SAR, China}

\author{K.-T.~Lin}
\thanks{These authors contributed equally}
\affiliation{Department of Physics and Center for Quantum Science and Engineering, National Taiwan University, Taipei 10617, Taiwan}

\author{F. Aziz}
\affiliation{Department of Physics, City University of Hong Kong, Kowloon, Hong Kong SAR, China}
\affiliation{Department of Physics, National Tsing Hua University, Hsinchu 30013, Taiwan}

\author{J.-C.~Liu}
\affiliation{Department of Electronic and Computer Engineering, The Hong Kong University of Science and Technology, Clear Water Bay, Kowloon, Hong Kong SAR, China}

\author{K.-V. Rangdhol}
\affiliation{Department of Physics, City University of Hong Kong, Kowloon, Hong Kong SAR, China}

\author{Y.-Y. Yeung}
\affiliation{Department of Physics, City University of Hong Kong, Kowloon, Hong Kong SAR, China}

\author{Sen~Yang}
\affiliation{Department of Physics and IAS Center of Quantum Technologies, The Hong Kong University of Science and Technology, Clear Water Bay, Kowloon, Hong Kong SAR, China}

\author{Qiming~Shao}
\affiliation{Department of Electronic and Computer Engineering, The Hong Kong University of Science and Technology, Clear Water Bay, Kowloon, Hong Kong SAR, China}

\author{Xin Wang}
\affiliation{Department of Physics, City University of Hong Kong, Kowloon, Hong Kong SAR, China}

\author{G.-D.~Lin}
\affiliation{Department of Physics and Center for Quantum Science and Engineering, National Taiwan University, Taipei 10617, Taiwan}
\affiliation{Physics Division, National Center for Theoretical Sciences, Taipei 10617, Taiwan}
\affiliation{Trapped-Ion Quantum Computing Laboratory, Hon Hai Research Institute, Taipei 11492, Taiwan}

\author{Franco Nori}
\affiliation{Center for Quantum computing and Cluster for Pioneering Research, RIKEN, Wakoshi, Saitama 351-0198, Japan}
\affiliation{\mbox{Physics Department, The University of Michigan, Ann Arbor, Michigan 48109-1040, United States}}

\author{I.-C.~Hoi}
\email[e-mail:]{iochoi@cityu.edu.hk}
\affiliation{Department of Physics, City University of Hong Kong, Kowloon, Hong Kong SAR, China}

\date{\today}

\maketitle
\tableofcontents

\section{Experimental setup}
\begin{figure}[t!]
\includegraphics[width=\linewidth]{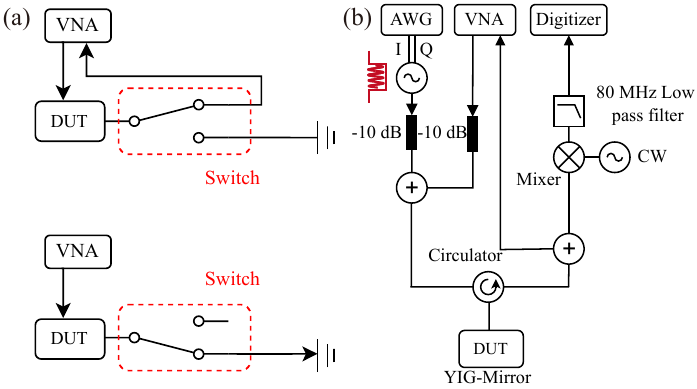}
\caption{
(a) Experimental setup for two distinct systems, separated by a microwave switch (red dashed box), without the mirror (top panel), and with the grounded mirror (bottom panel).
The device under test (DUT) is a YIG sphere placed on the coplanar waveguide (CPW). 
The CPW is fabricated on a $\unit[22\times50]{mm^2}$ RO4350B board, with a center conductor width of $\unit[1.1]{mm}$, a gap of $\unit[0.2]{mm}$ between the center conductor and the two ground planes, and a characteristic impedance of $Z_{0}\approx \unit[50]{\Omega}$. 
(b) Experimental setup  for both time- and frequency-domain measurements.
The DUT here is a YIG placed on the CPW at a distance $L_{4}=\unit[35.0]{cm}$ from the mirror. 
The vector RF source and AWG generate a microwave square pulse with carrier $\omega_{p}$, as indicated in the red rectangle. 
\label{fig:setup}}
\end{figure}

Figure~\ref{fig:setup}(a,b) illustrates the measurement procedures described in the main text.
The top and bottom panels in Fig.~\ref{fig:setup}(a) show the signal paths for the system without and with the mirror. 
A vector network analyzer (VNA) connected to both systems provides the input coherent continuous microwaves that interact with the Kittel
magnon mode of the YIG sphere, allowing us to measure the reflection and transmission coefficients ($S_{11}$
and $S_{21}$) to receive the YIG's response.
Figure~\ref{fig:setup}(b) shows that the measurement setup allows for both frequency- and time-domain experiments. 
In this setup, an arbitrary waveform generator (AWG) drives the in-phase-and-quadrature (IQ) modulated RF source. 
The digitizer at the end operates as a heterodyne receiver, processing the scattered fields of the YIG sphere after these pass through a down-converting
process (mixer and an RF source) and a low-pass filter.

\section{YIG-mirror distance calibration}
\begin{figure}[t!]
\includegraphics[width=\linewidth]{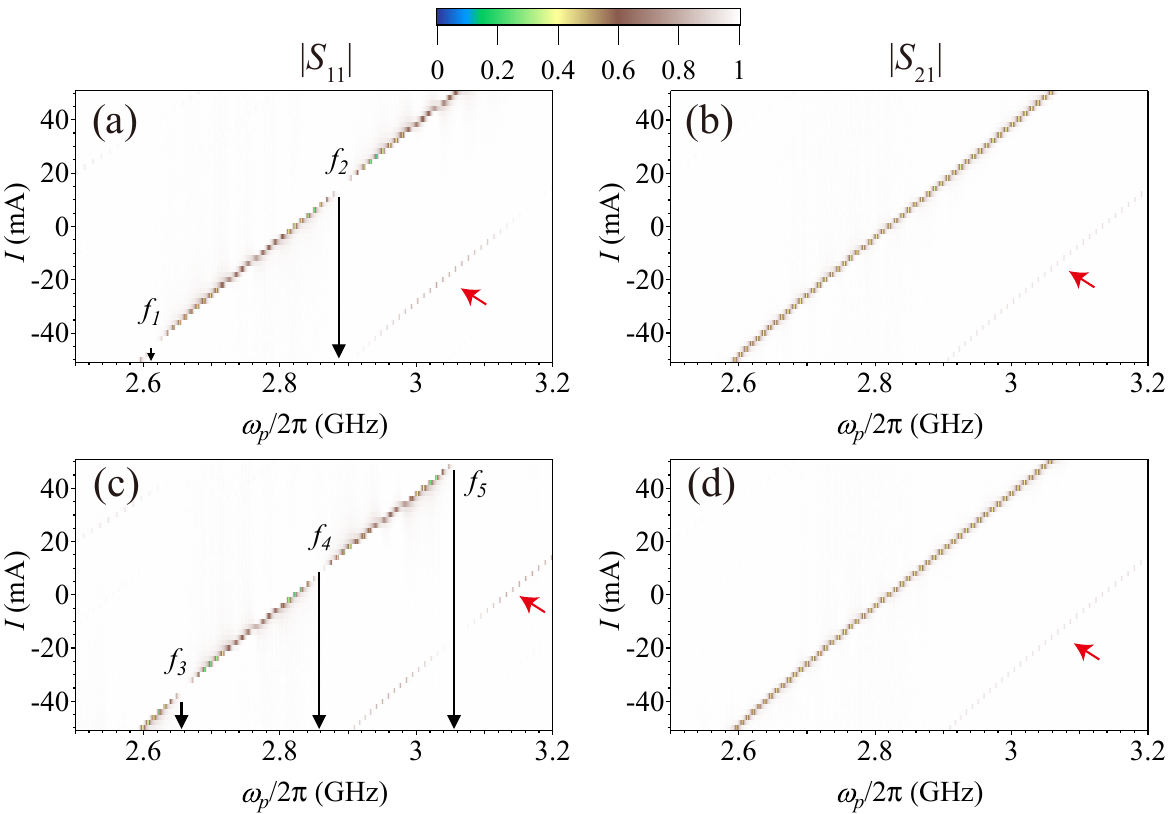}
\caption{
YIG sphere with $d=\unit[1.0]{mm}$ placed on top of the waveguide: (a) reflection coefficient at $L_{2}=\unit[38.1]{cm}$ away from the mirror and (b) transmission coefficient. 
(c) Reflection coefficient at $L_{3}=\unit[53.3]{cm}$ away from the mirror and (d) transmission coefficient.
The vertical black arrows denote the node frequencies ($f_1$, $f_2$, $f_3$, $f_4$, and $f_5$).
In addition to the Kittel mode, the red arrows indicate a high-order magnetostatic mode. 
\label{fig:1_0mm_38cm}}
\end{figure}

In this section, we aim to illustrate the calibration procedure for determining the distance between the YIG sphere and the mirror. 
In our setup, we use a commercial coaxial cable with speed of light $\nu\approx 0.7c$, and a length of $\unit[30.5]{cm}$ [from the switch to the ground mirror, see Fig.~1(b)]. 
The measured distance between the YIG sphere and the mirror is approximately $\unit[37\sim39]{cm}$.
To determine the actual distance $L_{2}$, we measure the reflection spectrum as shown in~\figpanel{fig:1_0mm_38cm}a and observe that the reflection magnitude appears periodically, compared to the transmission spectrum depicted in~\figpanel{fig:1_0mm_38cm}b.
The reflection magnitude $|S_{11}|=1$ occurs at the resonant frequency $f_{1}=\unit[2.6109]{GHz}$, and $f_{2} = \unit[2.8864]{GHz}$, indicating that the KM is at the magnetic nodes ($L_{2}/\lambda_{1}=4.75$ and $L_{2}/\lambda_{2}=5.25$, respectively) of the corresponding eigenmodes.
Thus the actual distance between the YIG sphere and the mirror is determined to be $L_{2}=4.75\nu/f_{1}=5.25\nu/f_{2}=\unit[38.1]{cm}$.
We conclude that $\bigtriangleup x=L_{2}-30.5=\unit[7.6]{cm}$ ($\bigtriangleup x$ remains unchanged when changing the cable lengths).

To further examine the speed of light $\nu$, we measure the reflection
and transmission spectra, as shown in~Figs.~\ref{fig:1_0mm_38cm}(c,d), with a cable
length of \unit[45.7]{cm}, corresponding to the total distance $L_{3}=45.7+7.6=\unit[53.3]{cm}$.
We observe that the magnetic nodes occur at frequencies $f_{3} = \unit[2.6558]{GHz}$, $f_{4} = \unit[2.8575]{GHz}$, and $f_{5}=\unit[3.0552]{GHz}$, enabling us to compute the corresponding wavelengths $\lambda=\nu/f=\unit[7.89]{cm},~\unit[7.34]{cm},~\unit[6.87]{cm}$, respectively.
Comparing the calculated $L_{3}/\lambda$ (6.755, 7.261, and 7.758) to the exact values (6.75, 7.25, 7.75) at these node frequencies, the differences are less than $0.1\%$. 
This verifies that the speed of light in this type of cable is $\nu=0.7c$.
With this calibration, we can extract the actual distance $L_{i}$ using the resonance frequency of KM at the node and the speed of light $\nu$.

\begin{figure}[t!]
\includegraphics[width=\linewidth]{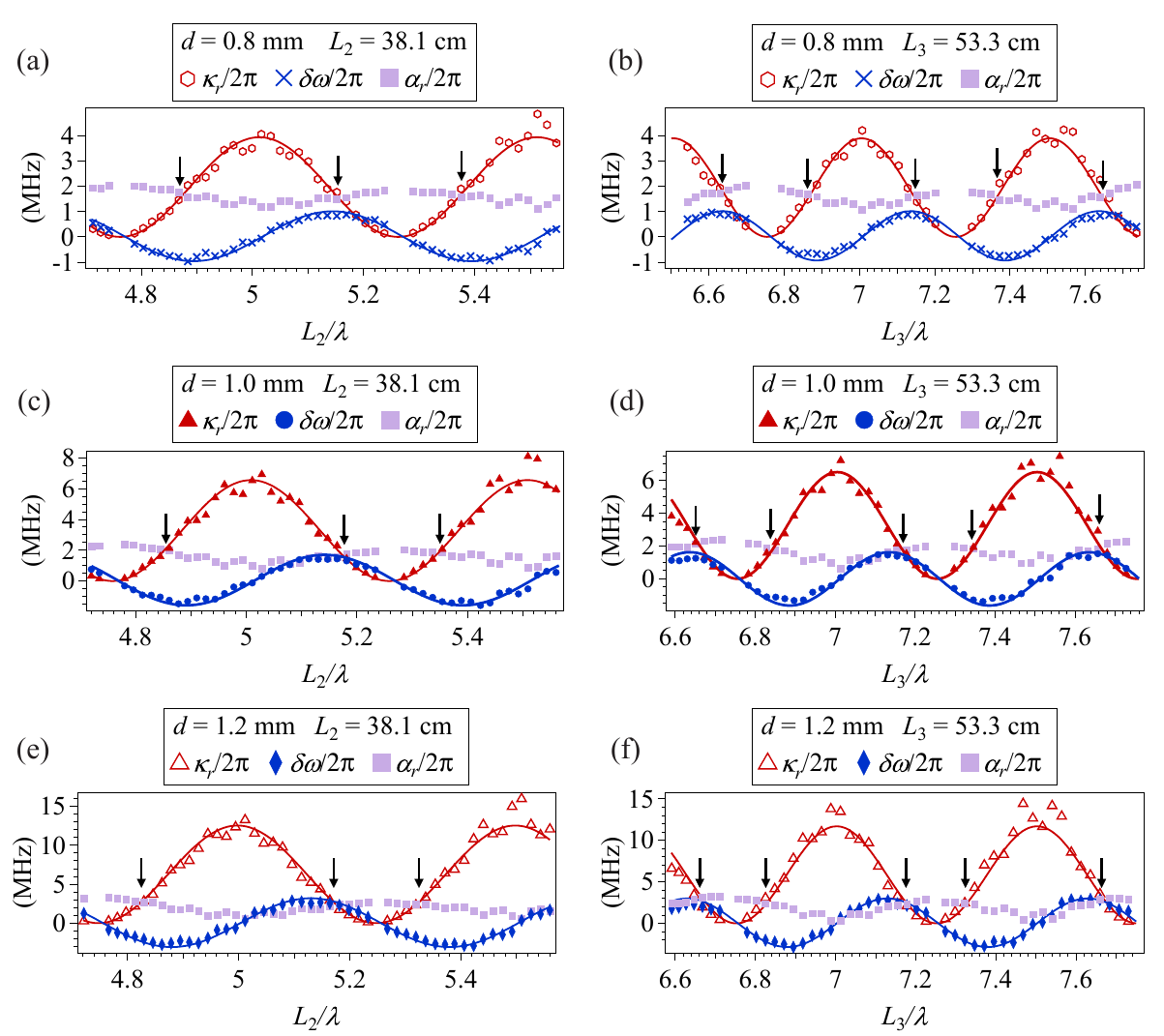}
\caption{ 
(a,c,e)  Radiative damping rate $\kappa_{r}$, the
resonant shift $\delta\omega$ and non-radiative damping rate $\alpha_{r}$ as functions of distance $L_{2}/\lambda$, corresponding to the YIG spheres with $d=\unit[0.8, 1.0,\rm and~1.2]{mm}$, respectively. 
Among them, in the mirror system, the distance $L_{2}$ is $\unit[38.1]{cm}$. 
(b,d,f) The difference from (a,c,e) is the distance from the mirror: $L_{3}=\unit[53.3]{cm}$.
Experimental data (symbols) are fitted by theoretical simulation (solid lines) based on Eqs.~(1,2,3) in the main text.
The black arrows show examples of critical coupling points, where $\kappa_{r}=\alpha_{r}$ at resonance.
}
\label{fig:summary}
\end{figure}
\begin{table}[H]
	\centering
	\begin{tabular}{| c | c | c | c |}
		\hline
		$\rm YIG$ &
		$L_{1} = \unit[22.8]{cm}$&
		$L_{2} = \unit[38.1]{cm}$&
		$L_{3} = \unit[53.3]{cm}$ \\
		\hline
		&
		$\kappa_{b}$~(MHz)&
		$\kappa_{b}$~(MHz)&
		$\kappa_{b}$~(MHz)\\
		\hline
		$\rm d=\unit[0.8]{mm}$&
		$1.877\pm0.030$ &
		$1.968\pm0.040$ &
		$1.946\pm0.052$ \\
		\hline
		$\rm d=\unit[1.0]{mm}$&
		$3.092\pm0.034$ &
		$3.280\pm0.044$ &
		$3.256\pm0.053$ \\
		\hline
		$\rm d=\unit[1.2]{mm}$&
		$5.845\pm0.030$ &
		$6.266\pm0.042$ &
		$6.202\pm0.052$ \\
		\hline
	\end{tabular}
	\caption{Summary of the bare radiative damping rates $\kappa_{b}$ of the Kittel mode for different YIG-mirror distances $L_{i}$ in \figref{fig:summary}, Fig.~2(e) and Fig.~3(a).}
	\label{tab:Parameters2}
\end{table}
\begin{figure}[t!]
\includegraphics[scale=0.5]{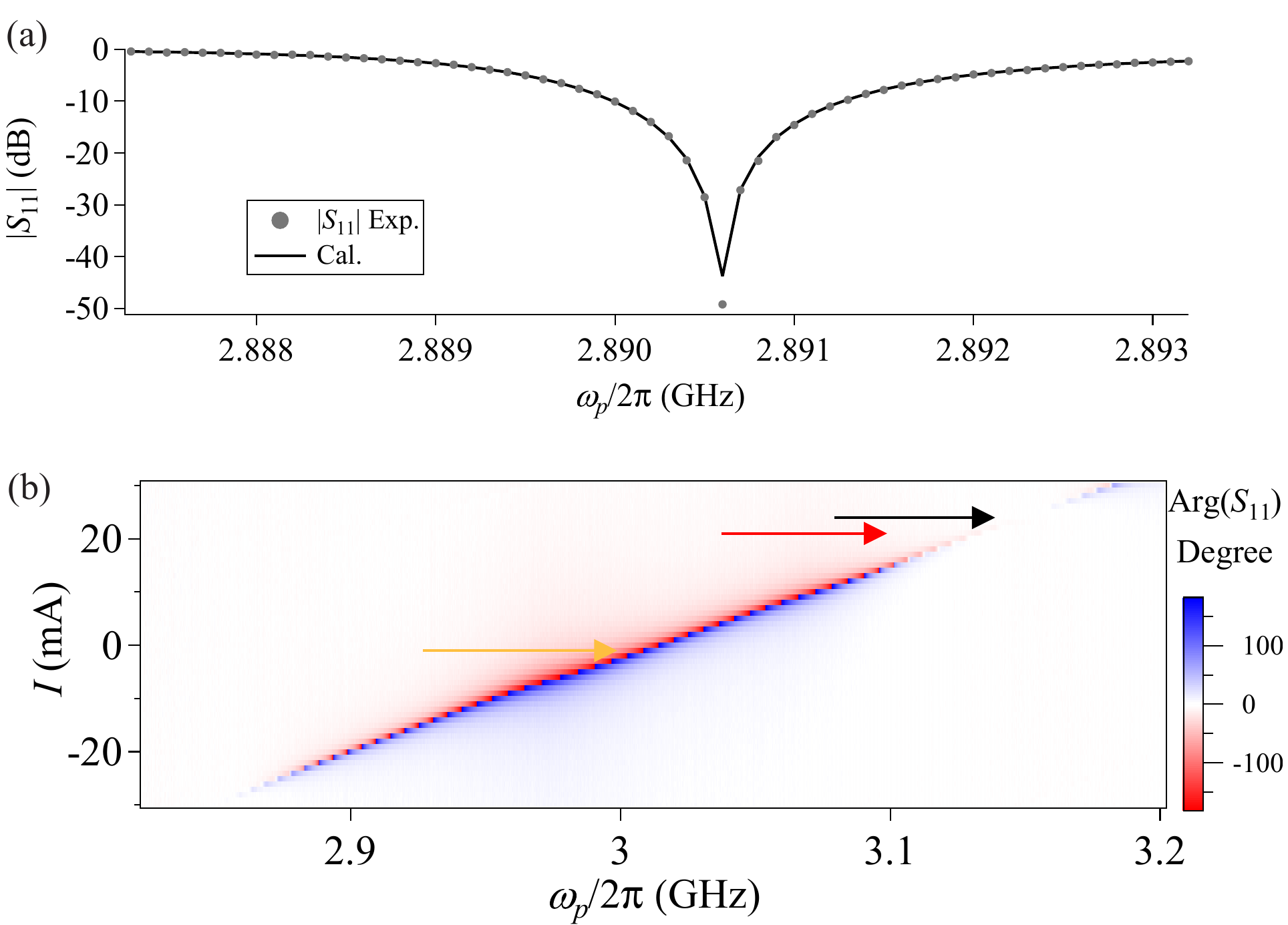}
\caption{
(a) The reflection magnitude $\left|S_{11}\right|$ of the YIG sphere with $d=\unit[0.8]{mm}$ with an ultra-high adsorption dip (\unit[$-$47.9]{dB}) at the critical coupling where the radiative damping rate $\kappa_{r}$ is equal to the non-radiative damping rate $\alpha_{r}$.
(b) 2D Phase response of the $S_{11}$ of the YIG as a function of the probe frequency $\omega_{p}$ and current $I$, with diameter $d=\unit[1.2]{mm}$. 
The Kittel mode resonance frequencies, indicated by the three arrows, are utilized for time-domain measurements.  
\label{fig:perfect_adsorption}}
\end{figure}

\section{Kittel mode in the frequency-domain measurements}
In this section, we summarize the frequency-domain measurement results. 
Figure~\ref{fig:summary} demonstrates the radiative damping rate $\kappa_{r}$, the
resonant shift $\delta\omega$ and non-radiative damping rate $\alpha_{r}$
against the distance $L_{i}$, based on the setup depicted in Fig.~1(b) and Fig.~\ref{fig:setup}(a).
We observe that these fundamental parameters exhibit a periodic structure, signaling the self-interference effect between the YIG and its mirror image.
Moreover, we notice that the extracted bare damping rate $\kappa_{b}$ is insensitive to the distance $L_{i}$, as shown in \tabref{tab:Parameters2}, which can be further confirmed using Eqs.~(1,2,3) in the main text. 
An intriguing observation is the increase and shift of critical coupling points when $\kappa_{r}=\alpha_{r}$.
In the case of $L_{2}=\unit[38.1]{cm}$, we find three interaction points, indicated by black arrows, while when $L_{3}=\unit[53.3]{cm}$, we have five interaction points.
At the critical coupling, we achieve an ultra-high adsorption dip in the reflection spectrum $\left|S_{11}\right|$, an example of which is shown in~\figpanel{fig:perfect_adsorption}{a}.

Switching to a different setup [refer to Fig.~\ref{fig:setup}(b)], we initially conduct frequency-domain measurements to ascertain the carrier frequencies $\omega_p$ of a series of pulses.
In ~\figpanel{fig:perfect_adsorption}{b}, indicated by the yellow arrow (together with the magnitude response), we perform a circle fit in the IQ plane to extract the parameters: $\omega_{m,r}/2\pi=\unit[3.008]{GHz}$, $\Gamma_{r}/2\pi=\unit[6.919]{MHz}$, and $\kappa_{r}/2\pi=\unit[11.57]{MHz}$.
However, at the node frequency marked by the black vertical arrow, the parameter extraction is not available due to an extremely weak KM response.
In this case, based on Eq.~(3) in the main text, we estimate that the KM is at the node, resonating at a frequency of $\unit[3.143]{GHz}$. 
We predict that the KM lifetime at the bias points, indicated by the yellow and red arrows, to be: $\tau=1/\Gamma_{r}=\unit[23.0, 90.6]{ns}$, respectively. 
Further details in the time-domain results are discussed in Section S4.

\section{Kittel mode in the time-domain measurements}
\begin{figure}[t!]
\includegraphics[width=\linewidth]{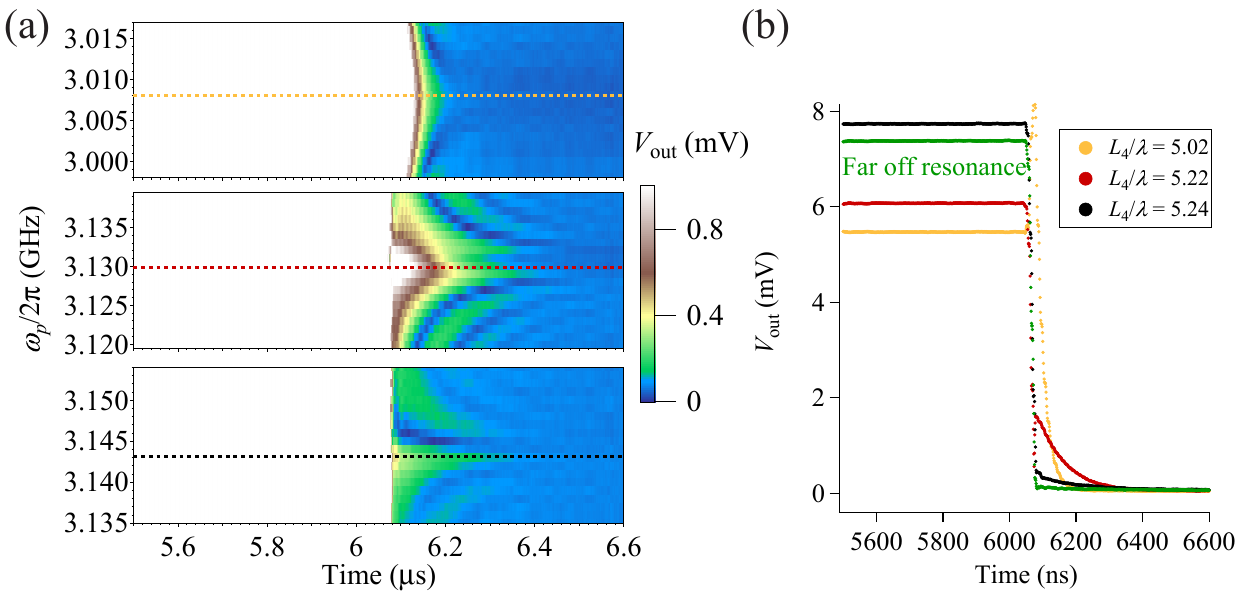}
\caption{(a) Emission of the Kittel mode versus carrier frequencies in the time domain at three bias points depicted in~\figpanel{fig:perfect_adsorption}{b}.
All the plots consists of two regions: the pulse-applied region on the left and the pulse off region on the right side.
The three dashed lines indicate that the square pulse resonates with the Kittel mode ($\omega_{p}=\omega_{m,r}$).
(b) The line-cuts of (a) are indicated by the dashed lines of the same color.  
\label{fig:time_domain}}
\end{figure}

In our time-domain experiment, we send rectangular microwave pulses with carrier frequencies $\omega_p$ to excite the KM [see \figpanel{fig:setup}b for setup].
To drive the KM to the steady state, the pulse duration is set to be much longer than the magnon lifetime 1/$\Gamma_{r}$.
To compare the spectroscopy results shown in ~\figpanel{fig:perfect_adsorption}{b}, the top to bottom panels in \figpanel{fig:time_domain}{a} correspond to three cases where the resonance frequencies of the KM are $\omega_{m,r}/2\pi=\unit[3.008]{GHz}$, $\unit[3.130]{GHz}$, and $\unit[3.143]{GHz}$. 
These settings place the YIG at normalized distances near the antinode ($L_{4}/\lambda=5.02$), between the antinode and node ($L_{4}/\lambda=5.22$), and near the node ($L_{4}/\lambda=5.24$), respectively. 

Figure~\ref{fig:time_domain}(b) shows the time dynamics of the KM when the frequency $\omega_{p}$ is resonant with the KM, as indicated
by the dotted lines in \figpanel{fig:time_domain}{a}. 
In the case where $L_{4}/\lambda=5.02$,
after turning off the probing pulse, we observe that the output voltage exhibits an overshooting peak, suggesting that \textit{the
energy stored in the YIG is coherently releasing back to the waveguide}.
Additionally, when the KM is tuned near the node, the magnon lifetime $\tau$ increases. 
To obtain $\tau$ for these three cases, we normalize the emission magnitude with $V_{\rm out}(t>t_{0})/V_{\rm out}(t_{0})$, where we set the maximum output $V_{\rm out}$ at the starting time $t_0=0$ [see Fig. 3(b) in the main text].
We fit the experimental data using an exponential function, yielding $\unit[23.1\pm0.1]{ns}$, $\unit[89.0\pm0.2]{ns}$, and $\unit[188.4\pm12.1]{ns}$, for the cases when $L_{4}/\lambda=$5.02, 5.22, and 5.24, respectively.

\begin{figure}[t!]
\includegraphics[width=\linewidth]{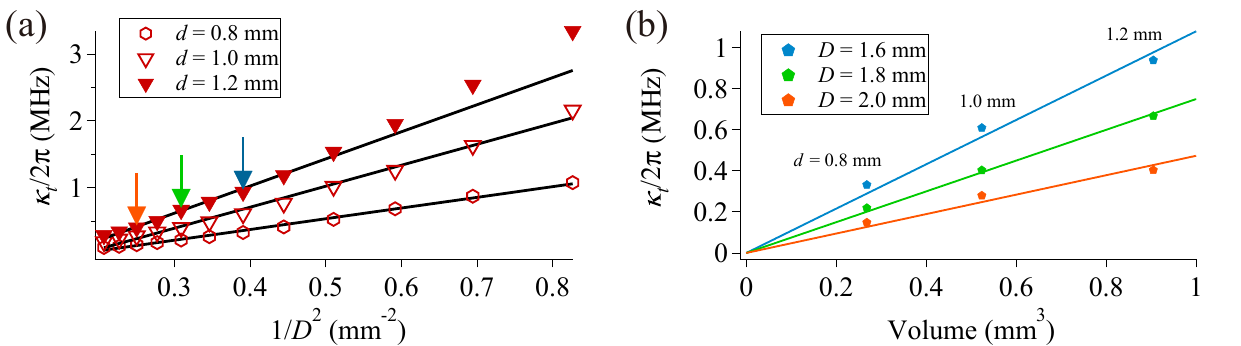}
\caption{(a) Radiative damping rate $\kappa_{t}$ of three different YIG spheres as a function of the inverse square of the height $1/D^{2}$, where the Kittel mode is at $\unit[2.6]{GHz}$.
(b) $\kappa_{t}$ as a function of volume for different height $D$, as indicated by the corresponding color arrows in (a).
\label{fig:spatial}}
\end{figure}
\section{Dependence of the collective coupling on both the height and the YIG sizes}
In this section, we study the effect of the YIG sphere diameter $d$ and its height $D$ above the CPW plane {[}see setup in Fig.~1(b){]} on the radiative decay $\kappa_{t}$ for the system without mirror. 
Figure~\ref{fig:spatial}(a) demonstrates the relation between $\kappa_{t}$ and $D$ for the case where $\omega_{m,t}/2\pi=\unit[2.6]{GHz}$.
We observe that, when the diameter $d=\unit[0.8]{mm}$, the damping rate $\kappa_{t}$ is linearly dependent on $1/D^{2}$, in agreement with Gauss's Law~\citep{Yang2018}.
However, when the diameter $d=\unit[1.2]{mm}$, this linear dependence breaks down, implying that the YIG sphere can no longer be treated as a point-like particle.
To further examine this effect, we perform an simulation on high frequency simulation software (not shown here) and find that there is a stronger field amplitude accumulated at the gap between the center conductor and the ground plane. 
This introduces a near-field effect, resulting in linewidth broadening~\citep{Yao2019} and large magnon-photon coupling~\cite{Rao_APL_2017,Yang2018}.

Moreover, we also demonstrate the YIG's size effect on $\kappa_{t}$, as shown in~\figpanel{fig:spatial}{b}.
We find that when $D$ is greater than $d$, $\kappa_{t}$ has a linear relationship with the volume of the YIG sphere, allowing us to relate the magnon-photon coupling strength $g$ to the number of spins $N$ as $g\propto\sqrt{N}$~\citep{Tabuchi2014, Justin2019}.

\section{Theory of magnons in a semi-infinite waveguide}

\subsection{Resonant dipole-dipole interactions}
In this section, we summarize the derivation of the effect of resonant
dipole-dipole interaction for the case where the KM
of a YIG sphere is coupled to a 1D semi-infinite
waveguide. The Hamiltonian describing the system is given by $H_{m}+H_{f}+V$
\citep{Wang2022,Zhan2022}, where $H_{m}=\hbar\omega_{m}a_{m}^{\dagger}a_{m}$
represents the energy of the magnons, and the field part is denoted by
$H_{f}=\hbar\int_{0}^{\infty}d\omega\omega b_{\omega}^{\dagger}b_{\omega}$.
In these expressions, $a_{m}$ ($a_{m}^{\dagger}$) represents the
annihilation (creation) of a magnon with energy $\hbar\omega_{m}$,
while $b_{\omega}^{\dagger}$ ($b_{\omega}$) is the creation (annihilation)
operator of a waveguide photon of frequency $\omega$.

The interaction Hamiltonian, 
\begin{equation}
V=\hbar\int_{0}^{\infty}d\omega g\left(\omega\right)\cos\left(k_{\omega}x\right)b_{\omega}^{\dagger}a_{m}+{\rm H.c.},\label{eq: interaction H}
\end{equation}
describes the interaction between a magnon located at position $x$
and a waveguide photon of frequency $\omega$. Here, the coupling
strength is modulated by $g\left(\omega\right)\cos\left(k_{\omega}x\right)$,
considering the boundary condition of an antinode at $x=0$ \citep{Lin2019,Wen2019,Nie_PRL_2023}.
In this expression, $k_{\omega}=\omega/\nu$ is the wavenumber with
speed of light $v$, and ${\rm H.c.}$ represents the Hermitian
conjugate. 

The dynamics of the field operator $b_{\omega}$ and the magnon operator
$a_{m}$ are described by the Heisenberg equations:
\begin{equation}
\frac{db_{\omega}}{dt}=-i\omega b_{\omega}-ig\left(\omega\right)\cos\left(k_{\omega}x\right)a_{m},\label{eq: field dynamics}
\end{equation}
\begin{equation}
\frac{da_{m}}{dt}=-i\omega_{m}a_{m}-i\int_{0}^{\infty}d\omega g\left(\omega\right)\cos\left(k_{\omega}x\right)b_{\omega}.\label{eq: magnon dynamics}
\end{equation}
Integrating Eq.~$\left(\ref{eq: field dynamics}\right)$ and applying
the Born-Markov approximation \citep{Lehmberg1970}, we obtain
\begin{equation}
b_{\omega}\left(t\right)=b_{\omega}\left(0\right)e^{-i\omega t}-\frac{g\left(\omega\right)\cos\left(k_{\omega}x\right)}{\left(\omega-\omega_{m}\right)-i\epsilon}a_{m}\left(t\right).\label{eq: b_omega_t}
\end{equation}
Here, the insertion of a small positive quantity $\epsilon$ ensures
the convergence of the integration. Substituting Eq.~$\left(\ref{eq: b_omega_t}\right)$
into Eq.~$\left(\ref{eq: magnon dynamics}\right)$, we obtain
\begin{equation}
\frac{da_{m}}{dt}=-i\sqrt{\kappa_{b}}\cos\left(k_{m}x\right)b_{in}\left(t\right)-i\omega_{m}a_{m}-iF\left(x\right)b_{\omega}\label{eq: magnon EoM}
\end{equation}
with the photonic operator $b_{\omega}\left(t\right)$ denoting the input
signal \citep{Scully_Zubairy_1997,Meystre2007,Agarwal_2012}, and 

\begin{equation}
F\left(x\right)=\int_{0}^{\infty}d\omega\frac{g^{2}\left(\omega\right)\cos^{2}\left(k_{\omega}x\right)}{\epsilon+i\left(\omega-\omega_{m}\right)}=\frac{\kappa_{r}}{2}+i\delta\omega,\label{eq: F function}
\end{equation}
accounting for the radiative damping rate $\kappa_{r}$ and resonance
shift $\delta\omega$. Using the technique of contour integration,
we obtain 
\begin{equation}
\kappa_{r}=2\kappa_{b}\cos^{2}\left(k_{\omega_{m}}x\right),\label{eq: decay rate}
\end{equation}
\begin{equation}
\delta\omega=\frac{\kappa_{b}}{2}\sin\left(2k_{\omega_{m}}x\right),\label{eq: energy shift}
\end{equation}
where $\kappa_{b}=\pi g^{2}\left(\omega_{m}\right)$ denotes the
bare damping rate in the 1D open waveguide. In the main text, 
\begin{equation}
\theta(I)=2k_{\omega_{m}}x,
\label{eq: phase_shift}
\end{equation}
where $k_{\omega_{m}}=2\pi/\lambda$ and $L_{i}=x$, as defined in Fig.~1 of the main text.
Equations~(\ref{eq: decay rate},\ref{eq: energy shift},\ref{eq: phase_shift}) recover Eqs.~(2,3,1) of the main text, respectively.

\subsection{Reflection coefficient $S_{11}$}

In order to probe the system, a weak coherent probe field with frequency
$\omega_{p}$ is introduced from the open end to interact with the
YIG sphere, as shown in Fig.~1(b). The corresponding reflection coefficient, measured at the same end, is determined by $S_{11}=\left\langle b_{\rm out}\left(t\right)\right\rangle /\left\langle b_{\rm in}\left(t\right)\right\rangle $.
Here, the photonic operator $b_{\rm out}\left(t\right)$, representing
the output signal, is connected to the input and the response of the YIG sphere through the input-output relation~\citep{Gardiner1985,Lalumiere2013}:
\begin{equation}
b_{\rm out}\left(t\right)=b_{\rm in}\left(t\right)-i\sqrt{\kappa_{b}}\cos\left(k_{\omega_{m}}x\right)a_{m}\left(t\right).\label{eq: input_ouput_relation_semi}
\end{equation}
Solving and substituting the steady-state solution of Eq.$\:$(\ref{eq: magnon dynamics})
into Eq.$\:$(\ref{eq: input_ouput_relation_semi}), the reflection
coefficient in the rotating frame with frequency $\omega_{p}$ is obtained:
\begin{equation}
S_{11}=1-\frac{\kappa_{r}}{\Gamma_{r}-i\left(\omega_{p}-\omega_{m}-\delta\omega\right)},\label{eq: reflection amplitude semi}
\end{equation}
where the overall decay rate is denoted by $\Gamma_{r}=\kappa_{r}/2+\alpha_{r}/2$.
The damping rate $\alpha_{r}$ is included to account for the non-radiative decay~\citep{Koshino_2012}.
In the main text, $\omega_{m,t}$ is equal to $\omega_{m}$, and $\omega_{m,r}=\omega_{m,t}+\delta\omega$.
Equation~(\ref{eq: reflection amplitude semi}) is shown as Eq.~(4) in the main text.

\section{Theory of magnons in open waveguides}

\subsection{Transmission coefficient $S_{21}$}

We now investigate the coupling of the Kittel mode of a YIG
sphere to a 1D open waveguide. A weak probe beam with frequency $\omega_{p}$
is introduced from the left-hand end to interact with the YIG system,
and the corresponding transmitted signal is then measured at the other end, as shown in Fig.~1(b). 
Our goal in this section is to derive the transmission amplitude $S_{21}$.

We start with the Hamiltonian $H=H_{m}+H_{f}+V$, which describes
this magnon-waveguide system. In this expression, the energy of the
magnons is denoted by $H_{m}=\hbar\omega_{m}a_{m}^{\dagger}a_{m}$,
and the energy of the fields is given by $H_{f}=\hbar\int_{0}^{\infty}d\omega\omega\left(b_{L}^{\dagger}\left(\omega\right)b_{L}\left(\omega\right)+b_{R}^{\dagger}\left(\omega\right)b_{R}\left(\omega\right)\right)$.
Here, the photonic operator $b_{L/R}\left(\omega\right)$ ($b_{L/R}^{\dagger}\left(\omega\right)$)
denotes the annihilation (creation) of a left-/right-propagating photon
with frequency $\omega$. The interaction part is given by 
\begin{equation}
V=\hbar\int_{0}^{\infty}d\omega a_{m}^{\dagger}\left(g_{L}\left(\omega\right)b_{L}\left(\omega\right)+g_{R}\left(\omega\right)b_{R}\left(\omega\right)\right)+{\rm H.c.},\label{eq: interaction V open}
\end{equation}
where ${\rm H.c.}$ represents the Hermitian conjugate and $g_{L/R}\left(\omega\right)$
describes the coupling strength between the magnon and the left-/right-going
photon of frequency $\omega$. Notably, in our experimental setup,
the coupling strength is expected to be symmetric: $g_{L}\left(\omega\right)=g_{R}\left(\omega\right)=g\left(\omega\right)$.
The Heisenberg equation of motion for the field and magnon operators
are given by
\begin{equation}
\frac{db_{R}\left(\omega,t\right)}{dt}=-i\omega b_{R}\left(\omega,t\right)-ig\left(\omega\right)a_{m}\left(t\right),\label{eq: dynamics bR}
\end{equation}
\begin{equation}
\frac{db_{L}\left(\omega,t\right)}{dt}=-i\omega b_{L}\left(\omega,t\right)-ig\left(\omega\right)a_{m}\left(t\right),\label{eq: dynamics bL}
\end{equation}
and 
\begin{equation}
\frac{da_{m}}{dt}=-i\omega_{m}a_{m}-ig\left(\omega\right)\int_{0}^{\infty}d\omega\left(b_{L}\left(\omega,t\right)+b_{R}\left(\omega,t\right)\right).\label{eq: dynamics am open}
\end{equation}

With the input-output relation discussed in the previous section,
the output signals $b_{out}^{L/R}\left(t\right)$ can be expressed
as 
\begin{equation}
b_{\rm out}^{R}\left(t\right)=b_{\rm in}^{R}\left(t\right)-i\sqrt{\kappa_{t}}a_{m}\left(t\right)\label{eq: input outout bR}
\end{equation}
and
\begin{equation}
b_{\rm out}^{L}\left(t\right)=b_{\rm in}^{L}\left(t\right)-i\sqrt{\kappa_{t}}a_{m}\left(t\right),\label{eq: input output bL}
\end{equation}
where $b_{\rm in}^{L/R}\left(t\right)$ represents the inputs, and the
radiative damping rate is denoted by 
\begin{equation}
\kappa_{t}=\pi g^{2}\left(\omega_{m}\right).\label{eq: coupling}
\end{equation}
Integrating Eq.$\:$(\ref{eq: dynamics bR}) and Eq.$\:$(\ref{eq: dynamics bL})
and substituting the corresponding solutions into Eq.$\:$(\ref{eq: dynamics am open}),
we arrive at 
\begin{equation}
\begin{aligned}\frac{da_{m}}{dt} & =-i\sqrt{\kappa_{t}}\left(b_{\rm in}^{L}\left(t\right)+b_{\rm in}^{R}\left(t\right)\right)-\left(\kappa_{t}+i\omega_{m}\right)a_{m},\end{aligned}
\label{eq: magnon EoM open case}
\end{equation}
whereby the transmission amplitude $S_{21}=b_{\rm out}^{R}\left(t\right)/b_{\rm in}^{R}\left(t\right)$
in the rotating frame of frequency $\omega_{p}$ is obtained as
\begin{equation}
S_{21}=1-\frac{\kappa_{t}/2}{\Gamma_{t}-i\left(\omega_{p}-\omega_{m}\right)},\label{eq: transmission amplitude}
\end{equation}
with $\Gamma_{t}=\kappa_{t}/2+\alpha_{t}/2$ being the overall damping
rate. 
Here, the damping rate $\alpha_{t}$ is incorporated to accommodate
the effects of non-radiative decay~\citep{Koshino_2012}.
Equation~(\ref{eq: transmission amplitude}) is shown as Eq.~(5) of the main text.


\bibliography{SI}